\documentclass[9pt,twocolumn,twoside]{osajnl}

\journal{jocn} 

\setboolean{shortarticle}{false}
\title{Large language models for optical network O\&M: Agent-embedded workflow for automation}

\author[1]{Shengnan Li}
\author[1]{Yidi Wang}
\author[1]{Fubin Wang}
\author[1]{Yujia Yang}
\author[1]{Yao Zhang}
\author[1]{Yuchen Song}
\author[2]{Xiaotian Jiang}
\author[3]{Yue Pang}
\author[1]{Min Zhang}
\author[1,*]{Danshi Wang}

\affil[1]{State Key Laboratory of Information Photonics and Optical Communications, Beijing University of Posts and Telecommunications, Beijing 100876, China}
\affil[2]{China Telecom Research Institute, Beijing, 102209, China}
\affil[3]{China Telecom Cloud Network Operating System R\&D Center, Beijing, 102299, China}
\affil[*]{danshi\_wang@bupt.edu.cn}

\begin{abstract}
With the continuous expansion of optical networks and the increasing diversity of services, existing operation and maintenance (O\&M) approaches are increasingly challenged to meet the rising demands for intelligence and efficiency. Large language models (LLMs), endowed with advanced semantic understanding and contextual analysis capabilities, are emerging as a promising enabler for intelligent optical network O\&M. Recent studies have demonstrated the feasibility of applying LLMs to optical network management, marking an important step toward intelligent automation. However, systematic investigations into how LLMs can be effectively integrated into existing O\&M workflows remain limited. This paper addresses this gap by drawing inspiration from best practices in real-world O\&M workflows and systematically identifying scenarios that are well suited for LLM integration. We highlight that agent-based design is key to improving the executability of tasks, and we propose a multi-Agent collaborative O\&M architecture that integrates LLM capabilities with existing O\&M tools. The proposed architecture leverages core LLM-related technologies including prompt engineering and tool invocation, to build Agent solutions targeting key tasks such as optical channel management, performance optimization, and fault management. This work presents a conceptual framework for embedding LLM-based Agents into optical network O\&M workflows, forming agentized processes that demonstrate the feasibility of LLM-assisted task execution and lay the groundwork for future autonomous O\&M systems featuring closed-loop perception, decision-making, and action.
\end{abstract}

\setboolean{displaycopyright}{false} % Do not include copyright or licensing information in submission.

\begin{document}

\maketitle

\section{Introduction}
Optical networks, with their advantages in high capacity, long-distance transmission, and strong reliability, play an irreplaceable role in high-speed data transport across regions and have become the cornerstone of modern information infrastructure. To meet the growing demands of emerging applications such as the Internet of Things (IoT), ultra-high-definition video, and intelligent computing, the capacity and scale of optical networks continue to expand rapidly. Ensuring high-reliability and lossless transmission in such large-scale networks requires highly efficient, flexible, and intelligent operations and maintenance (O\&M) capabilities \cite{tanaka2020autonomous}.

Currently, most telecom operators and cloud service providers rely on a network management system (NMS)-based O\&M framework \cite{tanaka2021monitoring}. Through NMS, a visualized topology of the deployed network can be constructed, control commands can be issued to optical network elements via integrated controllers, and device status and performance data can be efficiently retrieved. When optical network faults occur, related alarms are displayed in a centralized interface. This framework supports a variety of essential O\&M tasks, including power equalization, performance monitoring, and fault localization. However, today's O\&M processes still heavily depend on manual operations. For example, after alarms are triggered by device faults, operators are responsible for correlating alarm relationships, analyzing performance indicators via NMS, identifying root causes, and communicating with field engineers via text or voice to complete fault repairs. Similarly, when transmission capacity becomes insufficient, newly added channels are manually configured based on operator experience. To standardize manual operations and reduce human errors, detailed methods of procedure (MOP) and standard operating procedure (SOP) are typically followed during task execution. Nonetheless, with the rapid expansion of network scale and complexity, manual and human-based O\&M modes are facing growing limitations in efficiency and scalability. There is an urgent industry-wide need to enhance automation in O\&M processes.

Over the past decade, the application of artificial intelligence (AI) to optical network O\&M has attracted increasing attention \cite{musumeci2018overview}. Numerous studies have demonstrated that machine learning and deep learning techniques have strong potential in tasks such as fault prediction, anomaly detection, and quality of transmission (QoT) estimation \cite{khan2019optical,wang2022review,musumeci2025failure}. While effective within specific tasks, traditional AI models are typically designed for single-purpose applications and lack generalizability. They depend on manually defined input parameters and cannot perceive contextual information or interpret multi-round instructions. Moreover, they are incapable of performing task-level planning or tool orchestration, which limits their ability to achieve cross-task collaboration and autonomous decision-making. These limitations create a bottleneck for advancing the overall intelligence of network O\&M systems.

In recent years, large language models (LLMs) have made groundbreaking advances in natural language understanding, content generation, and natural language interaction. These capabilities offer new opportunities to build higher-level intelligent O\&M systems for optical networks \cite{wang2024large,sun2025experimental,liu2025first,sun2024first}. Notably, LLM-powered AI Agents are capable of interpreting complex natural language instructions, decomposing them into executable tasks, and autonomously invoking tools to complete operations, thereby eliminating the need for human-in-the-loop orchestration \cite{sun2024demonstration}. This opens up the potential for significantly improving automation in optical network operations \cite{di2024open,song2025synergistic,zhang2024gpt,zhang2025generative,zhang2025design}, positioning LLMs as a key enabling technology for future autonomous optical networks. A growing body of research has begun exploring the integration of LLMs into optical network O\&M scenarios, investigating their potential in tasks such as root cause alarm analysis, command generation, and configuration recommendation \cite{wang2023alarmgpt,wang2024alarmgpt,pang2024large,jiang2024opticomm,wang2025llm,wang2024llm,zhou2024large,zaid2025multi}. These studies provide early evidence of LLMs' ability to break away from human-dominated operational workflows. However, how to effectively leverage existing O\&M paradigms to improve operational efficiency when deploying LLMs at scale in optical networks has not been sufficiently explored.

This paper explores the application prospects of LLMs in large-scale optical network O\&M, with the goal of identifying representative use cases, outlining feasible architecture designs, and examining potential implementation pathways. Building upon existing operational frameworks, we advocate embedding LLM-based Agents into best-practice O\&M workflows, thereby forming Agent-embedded workflows that enhance automation and decision-making while maintaining compatibility with current operational paradigms. First, we provide a system-level review of key O\&M scenarios where LLMs can offer the most value, grounded in operational practices. We then propose a multi-Agent architecture tailored for optical channel provisioning, network performance optimization, and fault management, detailing the functional responsibilities and workflows of each Agent. Finally, the major challenges of deploying LLMs in real-world networks are discussed, including data availability, high-fidelity digital twin (DT) \cite{song2025lifecycle}, and security compliance. It is expected that this work can provide useful insights and a conceptual foundation for developing agentized workflows that bridge current O\&M practices with emerging LLM-driven automation.

The rest of the paper is organized as follows: Section 2 presents an overview of the physical, control, and application layers of optical networks and analyzes the pain points of current O\&M practices. Section 3 discusses the core LLM-related technologies that enhance automation in optical network operations. Section 4 introduces a multi-Agent architecture and details the design of intelligent Agents for key O\&M scenarios. Section 5 outlines the major deployment challenges of LLMs in real-world optical networks. Section 6 concludes the paper.

\section{Overview of Current Optical Network O\&M}

\begin{figure*}[htbp]
\centering
\includegraphics[width=0.77\textwidth]{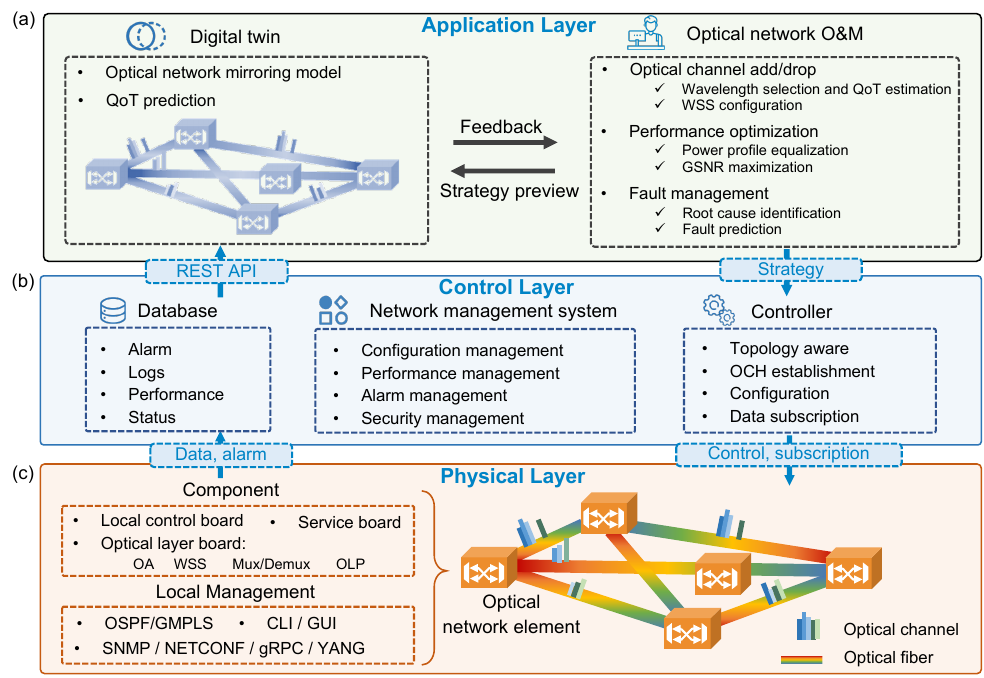}
\caption{Three-layer architecture of an optical network. (a) The physical layer comprises optical network elements and optical fibers, where each network element includes both hardware components and a local management system. (b) The control layer consists of the network management system (NMS), SDN controller, and databases. (c) The application layer encompasses the digital twin of the optical network and O\&M applications, including optical channel management, performance optimization, and fault management.}
\label{Fig1}
\end{figure*}

In this section, we presents an overview of the three-layer architecture of optical networks, followed by a discussion of three representative O\&M scenarios.

\subsection{Three-Layer Architecture of Optical Networks}

An optical network architecture typically consists of three layers: the physical layer, the control layer, and the application layer, as illustrated in Fig. \ref{Fig1}.

\textbf{In the physical layer:} optical network elements (NEs) are interconnected via optical fibers to form the network topology, as shown in Fig. \ref{Fig1}(c). The core hardware components of NE include the local control board, service board, and optical-layer board. The local control board communicates with other boards within the NE, issuing control commands and collecting monitoring data. The service board interfaces with switches and routers to inject service traffic into the optical network using standard optical wavelengths. The optical-layer board provides essential functions such as optical amplification, multiplexing/demultiplexing, optical switching, and optical protection, enabling long-haul, high-capacity transmission.

Given the typical lifecycle of an optical network exceeds 20 years, the network scale expands continuously over time. As a result, device from different vendors and technology generations often coexists within the same network, forming a heterogeneous and complex device environment. Additionally, the fiber link structure becomes increasingly intricate, with highly entangled service connectivity and diverse fault scenarios. To ensure manageability, each optical NE is equipped with a comprehensive local management system. These NEs communicate with the NMS via protocols such as OSPF, enabling centralized control. Local command-line interface (CLI) or graphical user interface (GUI) allows operators to configure device functionalities and query performance, alarms, and logs. In the event of NMS failure, local control capabilities are retained.

Simple network management protocol (SNMP) has long been the mainstream northbound management protocol, supporting efficient and reliable monitoring and configuration by the NMS. With the development of software-defined networking (SDN), northbound interfaces are evolving towards protocols such as NETCONF and gRPC \cite{martinez2025autonomous,manso2021tapi,giorgetti2020control}. Combined with YANG models, this enables standardized device modeling and fine-grained telemetry reporting at the second level or finer, facilitating precise network awareness and optimization \cite{miao2022detecting}.

\textbf{In the control layer:} the architecture typically consists of database, NMS, and controller, as illustrated in Fig. \ref{Fig1}(b). A centralized database ingests performance, status, alarm, and log data from physical-layer devices and stores information about network topology and service mappings. The database provides interfaces to the NMS, SDN controller, and DT. The NMS, as the traditional centralized platform, has long played a key role in unified device management, configuration provisioning, performance monitoring, and fault handling. It offers a unified graphical interface, particularly useful in multi-vendor and multi-technology environments. In recent years, SDN controllers have been introduced into the control layer and are gradually replacing parts of the NMS’s control functions. The goal is to support automated topology discovery, service deployment, and network optimization through centralized intelligence. Controllers typically obtain real-time telemetry data from devices via subscription mechanisms, thus creating a complete and dynamic network view that provides a solid data foundation for DTs.

In modern optical networks operated by cloud service providers, open and disaggregated architectures have been widely adopted \cite{xie2020open}. Vendor-developed SDN controllers are now capable of managing equipment from multiple suppliers, increasing deployment flexibility and, to some extent, reducing overall equipment costs \cite{le2022operator}. Driven by this trend, controller development has evolved beyond basic NMS-equivalent functions to support automatic topology construction and service provisioning, further expanding the capabilities of SDN-based network control.

\textbf{In the application layer:} DT is constructed to enable intelligent optical network O\&M, as depicted in Fig. \ref{Fig1}(a). DT technology, by accurately sensing physical-layer parameters and building high-fidelity mirror models, is expected to provide essential functions such as QoT estimation, optimization strategy verification, and fault analysis \cite{wang2024digital,zhuge2023building}. It has become a critical enabler for the intelligent evolution of optical networks. Recent research in both academia and industry has focused on improving the accuracy of DT models by calibrating the physical parameters of fibers and optical amplifiers. In current operational practices, DTs are primarily used for QoT estimation prior to configuration changes in live networks, effectively supporting operator decision-making and improving efficiency. Nevertheless, the end-to-end O\&M process of optical networks remains highly reliant on manual intervention and human coordination, limiting the level of automation. 

In general, optical network O\&M tasks can be categorized into three classes: optical channel management (i.e., add/drop), performance optimization, and fault management. The following section provide an overview of the current practices and main pain points of these three typical O\&M scenarios.

\subsection{O\&M Scenarios in Optical Networks}

\textbf{In optical channel management:} when increasing service demands require additional optical channels, operators must first evaluate resource availability between the relevant nodes. This is followed by routing and wavelength assignment (RWA) computation and QoT estimation to determine the configurations for service and optical-layer boards at each node. To minimize the impact on live traffic, such operations are typically conducted during off-peak night-time windows. Configuration commands are manually issued via the NMS, and the QoT of the new optical channels must be verified. Operators must also ensure that existing channels remain unaffected. The entire process typically takes several hours and is prone to human error. Optical channel deletion is considered a high-risk operation due to the possibility of accidental deletion of unrelated channels and is also scheduled during night-time maintenance windows.

\textbf{In performance optimization:} power equalization is one of the most common objectives. Optical power directly affects the optical signal-to-noise ratio (OSNR) and contributes to the accumulation of nonlinear Kerr effects, making it critical to transmission quality. Power flattening essentially ensures uniform performance across channels and is especially important in optical systems employing optical multiplex section protection (OMSP). When minor fiber degradation occurs without triggering protection switching, significantly lower power in one channel may reduce its OSNR below the threshold, causing persistent service disruption. In current operation practice, engineers typically perform power flattening checks periodically, often every few days in large-scale backbone networks, to ensure that all channels within an optical multiplex section (OMS) remain within a predefined flatness tolerance. Based on optical channel monitor (OCM) measurements, when deviations are detected, engineers manually adjust the variable optical attenuator (VOA) settings of wavelength selective switches (WSS)-equipped nodes, checking and correcting power levels segment by segment. For each wavelength, the adjustment and verification cycle generally takes several minutes, as engineers must coordinate between the NMS interface and on-site monitoring tools. For multi-span systems, downstream OMS segments often require additional corrections after upstream tuning, further increasing operational complexity. The process is labor-intensive, time-consuming, and vulnerable to misconfiguration that could interrupt services.

Even after power flattening is completed, the system may still not achieve optimal performance. A more advanced objective is to maximize the generalized signal-to-noise ratio (GSNR) of the optical transmission system. This process involves frequent interactions between GSNR optimization algorithms and the DT platform to identify the theoretical best configuration. Based on the feedback from live network performance after each configuration update, further parameter adjustments are made dynamically. This closed-loop optimization relies heavily on both model computation and real-time feedback, significantly increasing engineering complexity.

\textbf{In fault management:} the primary goal is to accurately locate the root cause of faults and take corrective actions as quickly as possible. In practice, a single fault can trigger a large number of correlated alarms, requiring operators to possess both deep domain knowledge and a comprehensive understanding of network topology. Research efforts in recent years have focused on alarm correlation and root cause analysis to enhance fault localization efficiency. LLMs have shown promise in extracting root causes from complex alarm patterns. During the recovery phase, field technicians and network operation engineers must collaborate for physical repairs. At present, fault information is often conveyed verbally, which increases the risk of communication and understanding errors that can delay repairs or even cause secondary fault. Another critical aspect of fault management is fault prediction. By detecting potential risks through predictive algorithms before a fault occurs, operators can proactively perform planned hardware replacements or implement other preventive measures to eliminate the root cause. In large-scale networks, timely and effective fault prediction requires the analysis of vast volumes of monitoring data in conjunction with the real-time interpretation of outputs from various predictive models. It is evident that such a task cannot be accomplished efficiently through manual efforts alone.

Once the optical network begins carrying live traffic, it enters a long-term O\&M phase that continues until decommissioning. This phase accounts for over 95\% of the network’s lifecycle, and its level of automation directly determines the overall intelligence of the network. However, current O\&M processes remain heavily dependent on manual intervention, resulting in inefficiencies and a high rate of human errors. These challenges make it increasingly difficult to cope with the rapid growth in network scale and complexity. To this end, introducing advanced decision-making mechanisms based on LLMs and efficiently leveraging existing AI algorithms for optical network operations has become a key pathway toward building a closed-loop intelligent O\&M system with sensing, decision, and execution capabilities. Therefore, the technique of AI is essential for advancing optical networks toward high-level autonomy.

%Example with the corresponding author designated by an asterisk:

%\author{Author One\authormark{1} and Author Two\authormark{2,*}}

%\address{\authormark{1}Peer Review, Publications Department,
%Optica Publishing Group, 2010 Massachusetts Avenue NW,
%Washington, DC 20036, USA\\
%\authormark{2}Publications Department, Optica Publishing Group,
%2010 Massachusetts Avenue NW, Washington, DC 20036, USA\\
%%\authormark{3}xyz@optica.org}

%\email{\authormark{*}opex@optica.org}}

% Example with the corresponding author designated by an asterisk and a note indicating equal contributions by two authors.

%\author{Author One\authormark{1,3} and Author %Two\authormark{2,3,*}}

% \address{\authormark{1}Peer Review, Publications Department,
% Optica Publishing Group, 2010 Massachusetts Avenue NW, %Washington, DC 20036, USA\\
% \authormark{2}Publications Department, Optica Publishing Group, %2010 Massachusetts Avenue NW, Washington, DC 20036, USA\\
% \authormark{3}The authors contributed equally to this work.\\
%\authormark{*}opex@optica.org}}

% \section{Examples of Article Components}
% \label{sec:examples}

\begin{figure*}[h]
\centering
\includegraphics[width=0.9\textwidth]{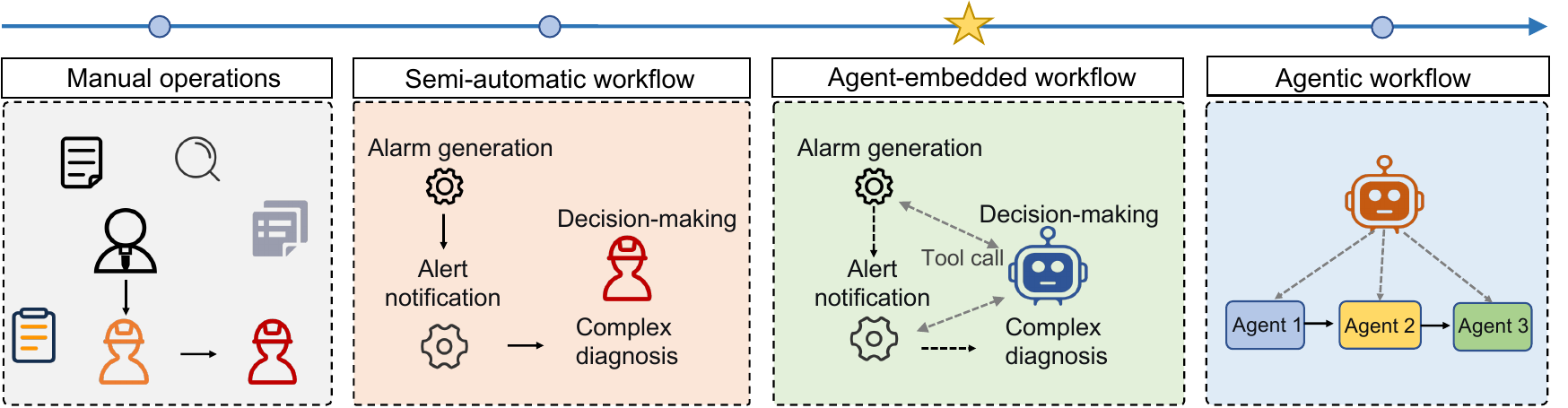}
\caption{Evolution of optical network O\&M workflows from manual operations to fully agentic workflows.}
\label{Fig2B}
\end{figure*}

\subsection{LLMs for Optical Network O\&M Automation}

The integration of AI technologies into optical networks has evolved in parallel with advances in the AI field itself. Traditional ML methods are lightweight and cost-effective to train, offering high precision and efficiency in solving well-defined network tasks. With the emergence of LLMs, which demonstrate exceptional capabilities in intent understanding, task decomposition, tool orchestration, and content generation, a promising path is emerging: building an autonomous operation system where LLMs serve as the intelligent control core, while conventional O\&M methods as robust execution tools. This architecture holds strong potential for enabling higher levels of automation in optical networks.

The application of LLMs in optical network automation is rapidly evolving. Existing studies have explored various high-value O\&M scenarios such as alarm analysis \cite{wang2023alarmgpt,wang2025graph}, log analysis \cite{pang2024large}, device configuration and performance optimization \cite{sun2025experimental,liu2025first,zaid2025multi}. To ensure sufficient contextual analysis and intent understanding capabilities, most studies adopt large-scale models from mainstream series such as GPT and LLaMA. Furthermore, \cite{sun2025experimental} addresses data security and privacy concerns by deploying local LLMs to build AI Agents capable of executing multiple O\&M tasks across the network lifecycle, providing a valuable example of matching model size with task complexity.

Building upon these initial explorations, the next step is to move beyond function-level demonstrations and explore how LLMs can be effectively embedded into existing O\&M workflows to enhance automation and decision-making. Instead of redesigning the O\&M framework from scratch, it is more feasible to build upon existing best-practice operational workflows and gradually evolve them toward LLM-driven architectures. In this way, LLMs can function as the central intelligence coordinating perception, analysis, and execution, ultimately improving operational efficiency and paving the way for more autonomous network management.

In the early stages of optical network O\&M, workflows were almost entirely manual, as illustrated in Fig. \ref{Fig2B}, where engineers had to consult documentation and communicate operational information through phone calls or other human-mediated channels. The current stage represents a semi-automated workflow, where the introduction of O\&M management platforms and analytical tools has significantly improved overall operational efficiency. This well-established workflow, refined through years of operational experience, ensures procedural standardization and system reliability. Therefore, when introducing Agents to further enhance O\&M efficiency, we advocate embedding them into the existing mature workflow rather than replacing it. The next stage of evolution will lead toward agentic workflows, where Agents autonomously orchestrate tasks and collaborate dynamically with other Agents based on network conditions. In this paper, our focus is on the Agent-embedded workflow, which serves as a key transitional step toward fully autonomous O\&M.

\begin{figure*}[htbp]
\centering
\includegraphics[width=0.95\textwidth]{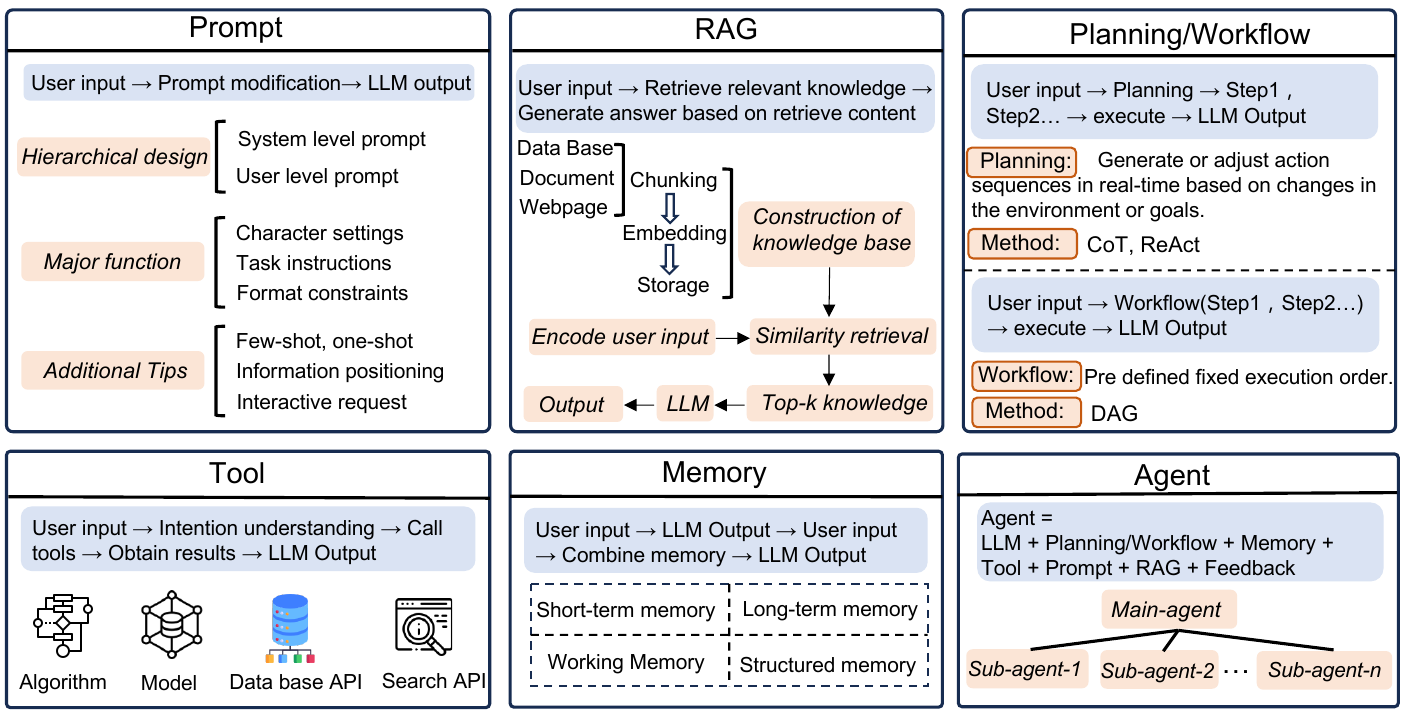}
\caption{Key enabling technologies for LLM-based solutions. The framework includes six core techniques—Prompt Engineering, RAG, Planning/Workflow, Tool Integration, and Memory—that collectively support the construction of robust and adaptable AI Agents in domain-specific scenarios.}
\label{Fig3}
\end{figure*}

\section{Key Technologies for LLM Deployment in Optical Networks}

With the rapid advancement of AI, LLMs have emerged as a significant milestone in natural language processing. Leveraging their powerful capabilities in text generation and understanding, LLMs are transforming how we think and solve complex problems. However, developing a optical network domain-adapted and reliable LLM-based solution remains a challenging task. From basic prompt engineering \cite{li2023prompt} to the orchestration of multi-Agent systems \cite{panait2005cooperative}, a variety of critical technologies are involved. This chapter provides an overview of these enabling technologies and clarifies their roles in supporting the integration of LLMs into optical network O\&M workflows.

Building an effective domain-specific LLM system requires efforts in two main directions: enhancing the base model’s capabilities, and developing robust supporting frameworks. The first direction includes fine-tuning techniques such as supervised fine-tuning (SFT) \cite{gunel2020supervised}, reinforcement learning with human feedback (RLHF) \cite{wang2024comprehensive}, and reinforcement fine-tuning (ReFT) \cite{trung2024reft}, as well as scaling model parameters and pretraining data. However, training or fine-tuning a domain-specific LLM from scratch is costly and often disproportionate to the practical benefits. By contrast, a more cost-effective approach is to build on top of powerful foundation models—such as the GPT series \cite{achiam2023gpt} or DeepSeek series \cite{liu2024deepseek}—through complementary techniques like prompt design and retrieval-augmented generation (RAG) \cite{jeong2023generative}. Therefore, in this context, we focus on how to adapt general-purpose LLMs to domain-specific tasks using auxiliary techniques, without modifying the underlying model parameters.

The development of an LLM-based system typically involves six core components: prompt design, RAG, planning and workflow, tool integration, memory mechanisms, and Agent construction, as illustrated in Fig. \ref{Fig3}. These components are illustrated in detail below.

Prompting is one of the earliest and most widely used techniques for improving LLM performance, which involves crafting input instructions to guide the model toward generating task-specific outputs. Prompts can be categorized into system-level and user-level prompts. The former defines global behavioral norms that persist throughout the session, while the latter is task-specific and often highly specialized. Functionally, a well-designed prompt typically includes three elements: role assignment, task instruction, and output formatting. Role assignment activates the model's domain knowledge (e.g., "You are a network operations expert"), often by leveraging patterns seen during pretraining. Task instructions provide a detailed description of the objectives and required operations, and formatting constraints ensure outputs meet structural expectations. Additionally, prompts may include optional tips such as few-shot examples, information localization hints, or interaction strategies (e.g., "Ask the user for clarification if needed") to improve robustness and interactivity.

RAG is one of the most effective methods for extending an LLM’s domain-specific knowledge. Its core idea is to retrieve relevant information from external sources and condition the model’s response on that content. The process involves constructing a knowledge base by chunking, embedding, and vectorizing documents (from databases, files, web content, etc.). When a user query arrives, it is also embedded and compared with the knowledge base to retrieve the top-k relevant entries \cite{wang2023alarmgpt}, which are then inserted into the prompt to guide the LLM’s response.

Planning \cite{huang2024understanding} and workflow \cite{grunde2023designing} are two key methods for organizing complex tasks. Planning emphasizes the model’s structured reasoning process, involving task decomposition, intent interpretation, and action sequencing. To enhance this capability, strategies such as Chain-of-Thought (CoT) prompting \cite{wei2022chain} and ReAct \cite{aksitov2023rest} are often employed to guide the model through intermediate reasoning steps in a more interpretable and reliable manner. In contrast, workflows are deterministic and structured, defined in advance as directed acyclic graphs (DAGs) \cite{digitale2022tutorial} to ensure stability and correctness. While planning offers flexibility and adaptability, workflows provide robustness for well-defined and repetitive tasks.

While LLMs are primarily language models, many tasks require interacting with external systems. Tools provide the LLM with operational capabilities \cite{shen2024llm}, functioning as the “hands and feet” of the model. These tools include algorithms, APIs (for databases, search engines, network management systems), and domain-specific plugins, enabling the LLM to execute real-world actions beyond text generation.

Memory mechanisms is essential for maintaining context in multi-turn interactions \cite{hatalis2023memory}. Depending on the design, memory may include short-term memory such as recent inputs, long-term memory that stores persistent knowledge, working memory used for task-specific context, and structured memory that contains organized representations. Memory systems are critical for coherent reasoning and task continuity.

The Agent is a central abstraction that encapsulates the aforementioned technologies into a coherent unit. An Agent may include prompts, tools, RAG modules, planning workflows, and memory. Agents are typically divided into two roles: a supervisor Agent that manages user interactions and task delegation, and sub-Agents that execute specific domain tasks. Each sub-Agent integrates its own tools and workflows to carry out specialized responsibilities.

\begin{figure}[h]
\centering
\includegraphics[width=\columnwidth]{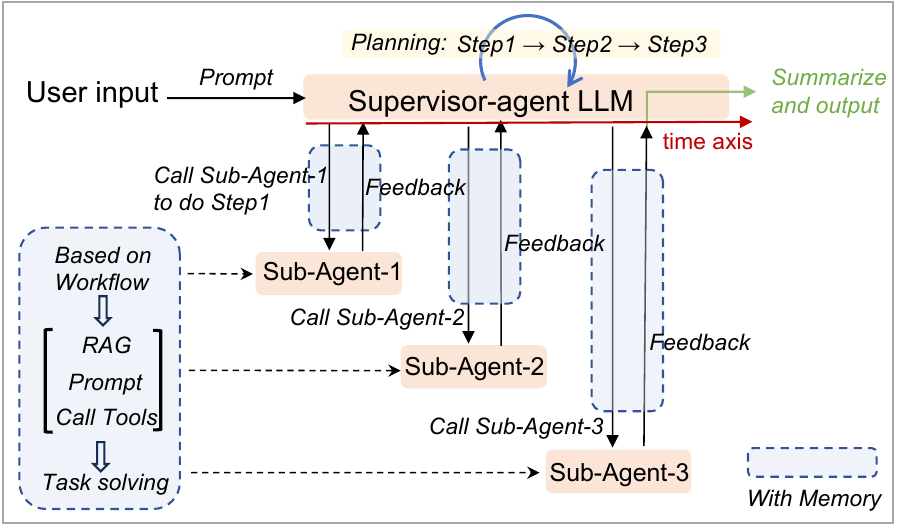}
\caption{Execution workflow of an LLM-based multi-Agent system. The Supervisor Agent interprets the user intent and coordinates subtasks among multiple specialized Sub-Agents, each utilizing tools such as prompt templates, RAG, and external APIs. The process iteratively refines the response until the final result is delivered.}
\label{Fig4}
\end{figure}

With all these components, an LLM-based system can achieve end-to-end task automation. The runtime process unfolds as follows: user input is first processed with prompts and sent to the Supervisor Agent, which interprets intent and decomposes the task into sub-tasks, as illustrated in Fig. \ref{Fig4}. It then dispatches each task to the corresponding sub-Agent in sequence. Each sub-Agent leverages its internal prompt, tools, and knowledge to generate an answer. If the result is satisfactory, it proceeds to the next task; otherwise, it revises or retries. After all tasks are completed, the Supervisor Agent compiles the results into a final output.

The technologies introduced in this chapter form a general-purpose, multi-Agent LLM framework. In the next chapter, we present example applications of this architecture in optical network operation scenarios, highlighting its potential use in automating key O\&M tasks

\section{LLM-Based AI-Agent for Optical Network Operation and Maintenance}

\begin{figure*}[htbp]
\centering
\includegraphics[width=0.95\textwidth]{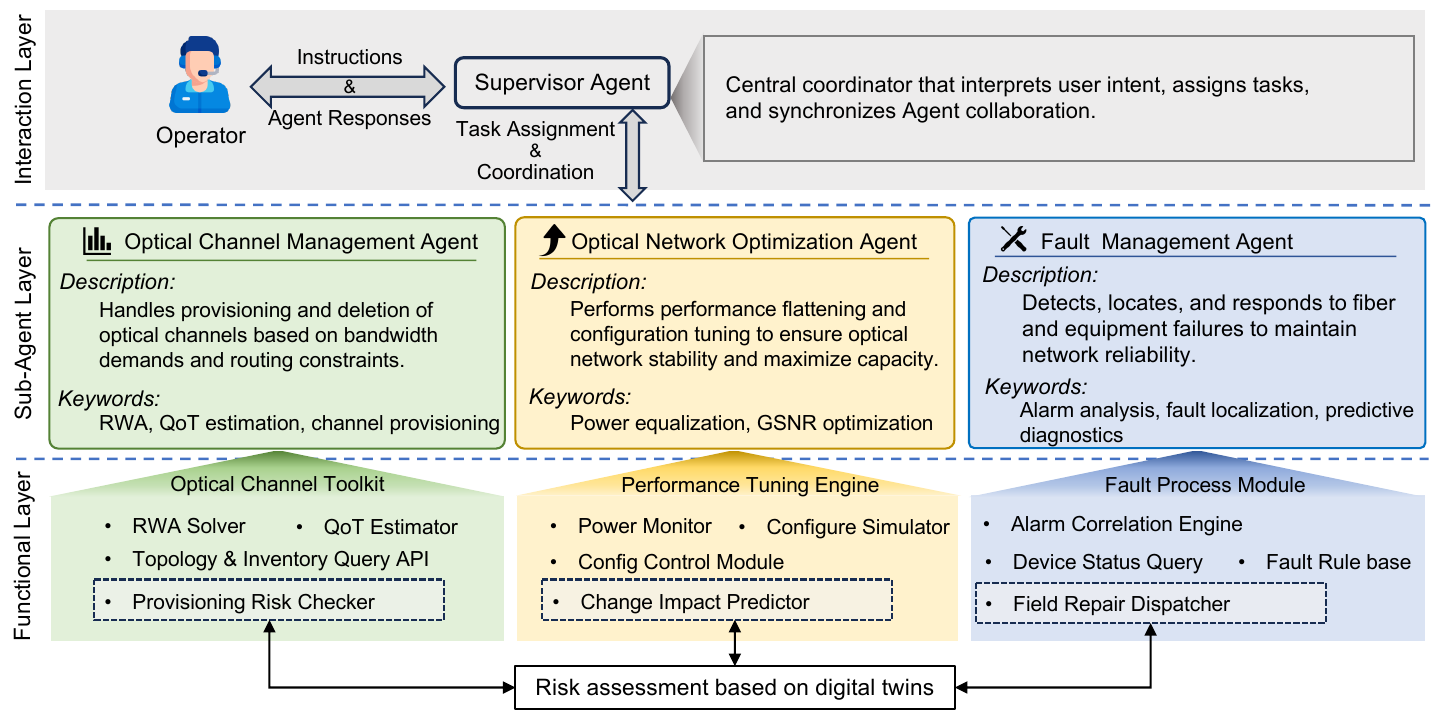}
\caption{Multi-Agent collaborative architecture for optical network operations and maintenance. The interaction layer includes bidirectional communication between human operators and the Supervisor Agent, as well as between the Supervisor Agent and task-specific Sub-Agents. The Sub-Agent layer includes three sub-Agents, each responsible for a key O\&M domain: optical channel management, performance optimization, and fault management. The function layer defines core capabilities required to fulfill each Sub-Agent’s responsibilities.}
\label{Fig5}
\end{figure*}

Although O\&M tasks in optical networks can be broadly categorized into three typical classes: optical channel management, performance optimization, and fault management, each category encompasses a wide range of complex sub-tasks. To simplify Agent design and improve task execution efficiency, it is necessary to assign dedicated sub-Agents to each task category. Moreover, since these tasks are often interrelated in practice (e.g., channel addition typically requires subsequent performance tuning), it is essential to establish coordination mechanisms among sub-Agents.

Given that multiple sub-Agents may act upon the same optical network simultaneously, a centralized coordination mechanism is required to prevent operational conflicts and ensure global network stability. To this end, a Supervisor Agent is introduced to orchestrate the execution of all sub-Agents and serve as the unified interface for user interaction. Based on the actual workflow of O\&M in current networks, we propose a multi-Agent architecture tailored to diverse O\&M scenarios.

\subsection{Multi-Agent Architecture for Optical Network O\&M}
The proposed multi-Agent cooperative O\&M framework is structured into three logical layers: the Interaction Layer, the Sub-Agent Layer, and the Functional Layer, as shown in Fig. \ref{Fig5}.

The Interaction Layer includes interactions between the Supervisor Agent and both human operators and sub-Agents. Although the goal of introducing LLMs and Agent-based systems is to reduce human involvement by enabling Agents to autonomously acquire network data and execute tasks, human operators still play an important role during the early stages of network automation. For example, optical channel add/drop operations are often driven by service-layer bandwidth demands. Since such demands may not yet be automatically communicated from the service side, or due to design choices to decouple IP and optical-layer intelligence, optical network engineers must manually specify channel-related operations to the Supervisor Agent. Furthermore, the Supervisor Agent is responsible for interpreting operator intent, decomposing tasks, and dispatching them to the appropriate sub-Agents. It also manages inter-Agent coordination when one Agent’s task requires collaboration from another.

The Sub-Agent Layer consists of three dedicated sub-Agents responsible for optical channel management, performance optimization, and fault management. Sub-Agents can either operate continuously following predefined workflows or respond to specific instructions issued by the Supervisor Agent. For security and control integrity, direct communication between sub-Agents is disallowed. Instead, coordination is exclusively handled by the Supervisor Agent. Each sub-Agent is responsible for defining its task boundaries, building dedicated workflows, and preparing the tools necessary to carry out its assigned functions. The scope and complexity of each sub-Agent’s tasks should align with the current level of automation in the network, taking into account factors such as digital twin modeling fidelity and network data accessibility.

The Functional Layer defines the essential capabilities required by each sub-Agent to complete its designated tasks. For example, the fault management Agent must support core functions such as root cause identification and the issuance of repair instructions to field engineers. More advanced functions, such as fault prediction or anomaly detection, can be introduced progressively based on the network's maturity in automation. Additionally, as optical equipment evolves (e.g., the deployment of new transceiver modules or board types), the capabilities of sub-Agents must also be updated to support new operational requirements.

In summary, the proposed multi-Agent architecture establishes two key design principles for the deployment of LLM-based AI Agents in optical network O\&M scenarios: first, assign dedicated sub-Agents for each major O\&M task category; second, implement a Supervisor Agent for unified coordination to ensure coherent and orderly automation. The proposed framework follows a centralized network management architecture, where the Supervisor Agent and sub-Agents are deployed close to the NMS servers to minimize data retrieval and command execution latency. In the following sections, we detail the design and implementation strategies for each of the three sub-Agents.

\begin{figure*}[htbp]
\centering
\includegraphics[width=0.99\textwidth]{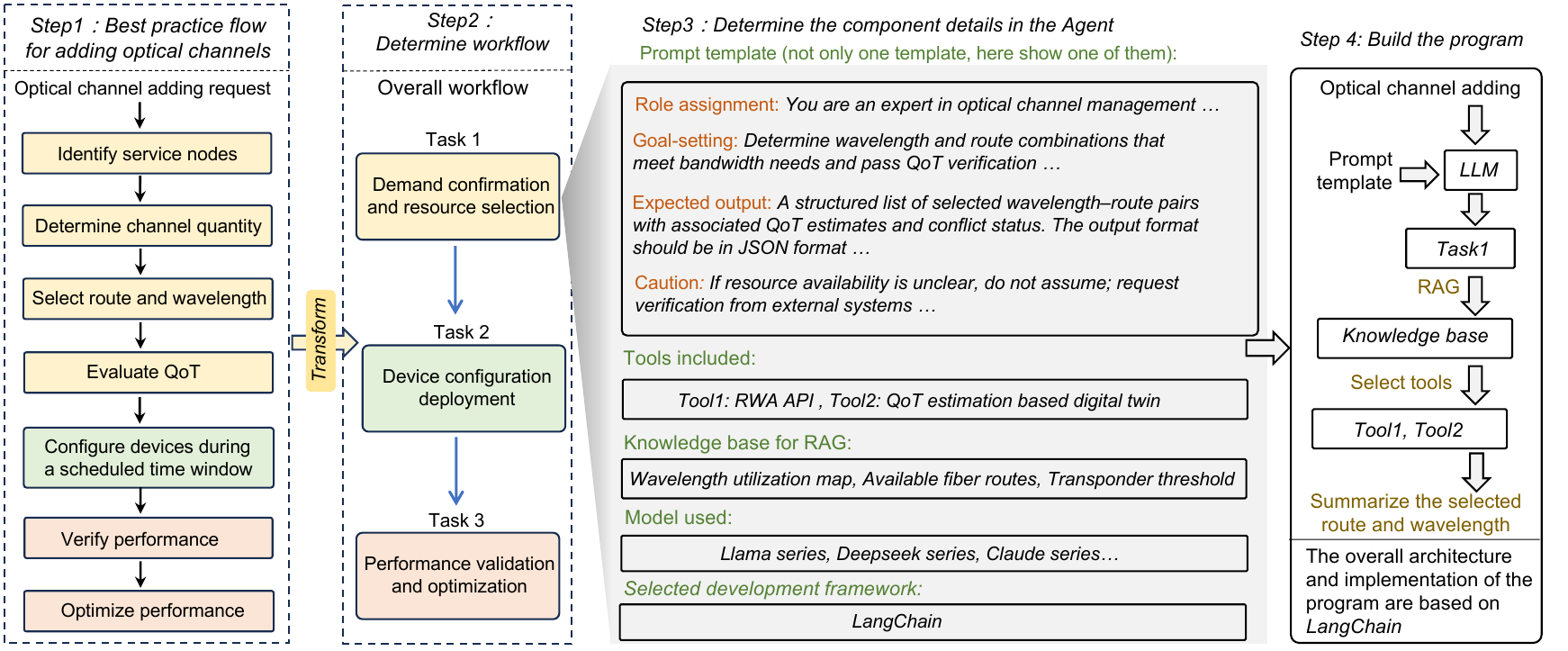}
\caption{Implementation framework of the optical channel management Agent for bandwidth-driven wavelength provisioning. The figure illustrates the practical workflow from operator requests to optical channel deployment, including real-world practices, multi-step Agent workflow, modular prompt-tool-RAG design, and standardized Agent development based on LangChain.}
\label{Fig6}
\end{figure*}

\subsection{Agent for Optical Channel Management}

In optical networks, the addition and deletion of optical channels generally arise from two scenarios: (1) changes in bandwidth demand between service nodes, and (2) wavelength re-routing that necessitates the reallocation of channels in certain OMS. The following discussion focuses on the first case, adding new optical channels to accommodate increased bandwidth requirements between two nodes, and presents the design methodology of the optical channel management sub-Agent.

Since LLM-based Agents possess strong intent understanding and, when augmented with external tools and knowledge bases, can execute tasks efficiently, it is important to first map out the current best-practice workflow followed by human engineers. In a typical add-channel scenario, the process begins with identifying the two service nodes between which additional bandwidth is needed and determining the number of optical channels required. RWA algorithm is then used to select the appropriate routes and spectrum slots. To ensure that transmission quality meets margin requirements, the selected spectrum paths must be validated using QoT estimation tools. Next, during a designated maintenance window, the network operator configures service boards and WSS modules to establish the new channels and verifies their performance. To minimize the risk of errors, these operations are usually atomized, with channels added in batches and optical power gradually tuned. If the resulting performance or power flatness is suboptimal, further performance optimization is needed, which will be detailed in the next section.

Once the best-practice workflow is clarified, the next step is to translate it into an executable workflow for the Agent, as shown in Fig. \ref{Fig6}. This automated workflow should mirror the manual process while segmenting tasks into manageable and logically grouped task nodes. In the channel addition scenario, the workflow can be divided into three main task nodes: (1) demand confirmation and resource selection, (2) device configuration deployment, and (3) performance validation and optimization. Each node consists of multiple sub-steps such as information retrieval or algorithm invocation. These nodes are executed sequentially, triggered either by the completion of the previous node or by explicit commands from the Supervisor Agent. For example, while demand confirmation can be executed anytime, device configuration must wait until an authorized maintenance window. The granularity of task nodes should remain flexible based on network complexity, tool maturity, and other constraints. The core principle is to maintain tight cohesion within each node while keeping the complexity manageable.

The next critical step before Agent implementation is the preparation of prompt templates for each task node, the selection of necessary tools and external knowledge bases, and the identification of suitable LLMs and development frameworks. We now take the first task node, demand confirmation and resource selection, as an example to explain this process.

When additional channels are required, the operator provides input to the Supervisor Agent using natural language. The Supervisor Agent interprets the request and passes key information to the optical channel management sub-Agent. This sub-Agent must extract relevant details such as node names and the number of required channels for use in subsequent routing and wavelength selection. For this purpose, two separate prompt templates can be created: one for demand confirmation and one for resource selection. In both cases, the LLM should be assigned the role of an optical channel management expert (e.g., "You are an expert in optical network channel management") to activate domain-specific knowledge and enhance understanding of technical terms and procedures.

In the demand confirmation prompt, the goal is to extract the service nodes and the required number of channels from the Supervisor Agent’s message and output them in a standardized format. This step does not require external tools or RAG, as it relies entirely on the LLM's text processing capability. The output is then passed as input to the resource selection step.

In the resource selection prompt, the role remains the same, but the goal shifts to identifying available spectrum and routing options. The prompt should include example output formats and instructions to flag uncertainties, minimizing the risk of hallucination. This step typically requires the invocation of an RWA algorithm to identify non-conflicting slots and paths, and the use of a DT interface to evaluate QoT for the selected spectrum to ensure minimal impact on existing channels. The knowledge base should include detailed information on wavelength usage, fiber routing lengths and health scores, as well as the OSNR thresholds required by OTUs under different modulation formats. Overall, this task node relies on the LLM’s ability to process textual inputs and analyze the results from RWA and DT tools, requiring a relatively strong general-purpose LLM. The final model choice should depend on factors such as network scale, service complexity, and empirical performance. LangChain is a recommended development framework for implementing multi-step workflows in a modular fashion. LangChain provides an abstraction layer for tool invocation. Within LangChain, the LLM automatically generates a structured call (e.g., a JSON object) following the predefined schema of available tools. The execution layer, rather than the LLM itself, runs the corresponding Python function or API, and the result is then automatically returned to the LLM for further planning and synthesis.

\begin{figure}[h]
\centering
\includegraphics[width=\columnwidth]{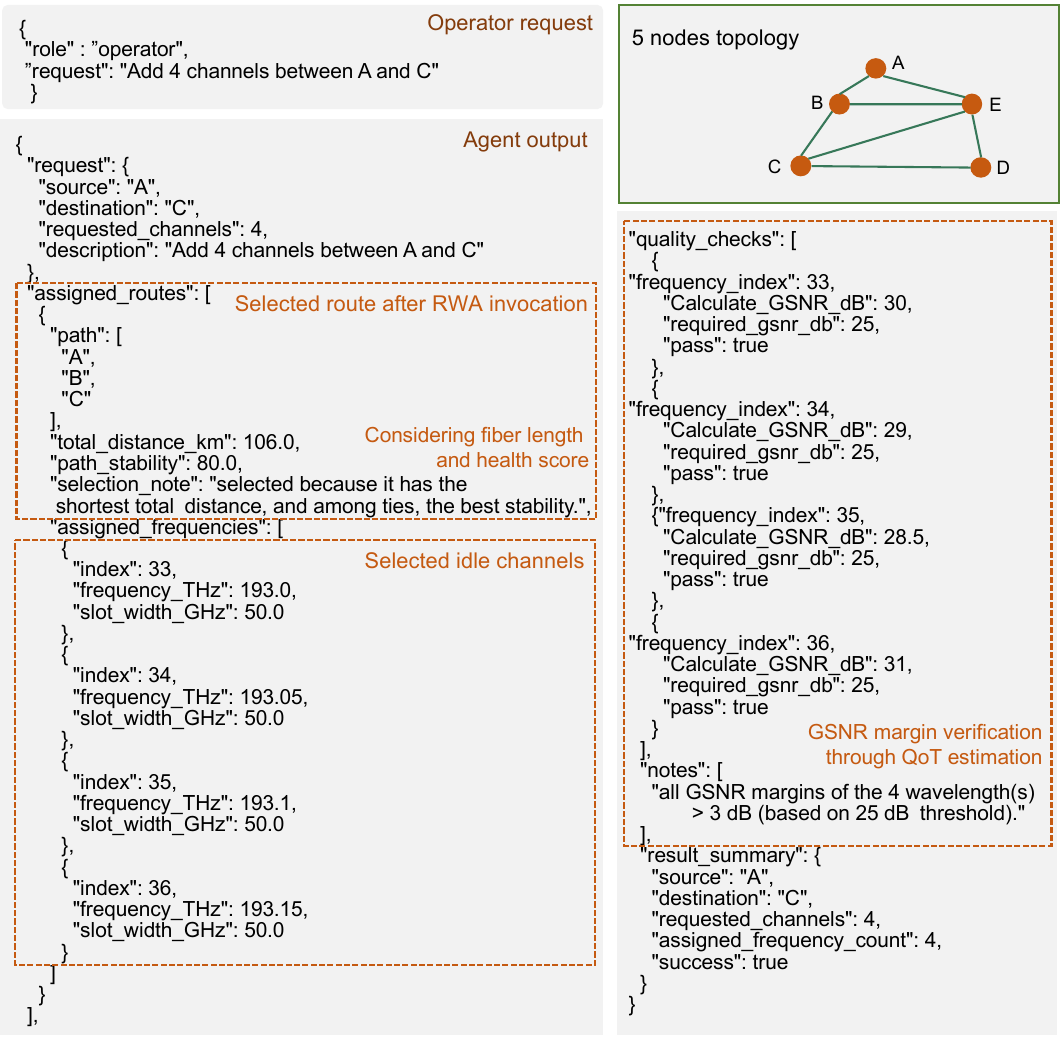}
\caption{Five nodes topology and illustrative validation results for demand confirmation and resource selection.}
\label{Fig6B}
\end{figure}

\begin{figure*}[htbp]
\centering
\includegraphics[width=0.99\textwidth]{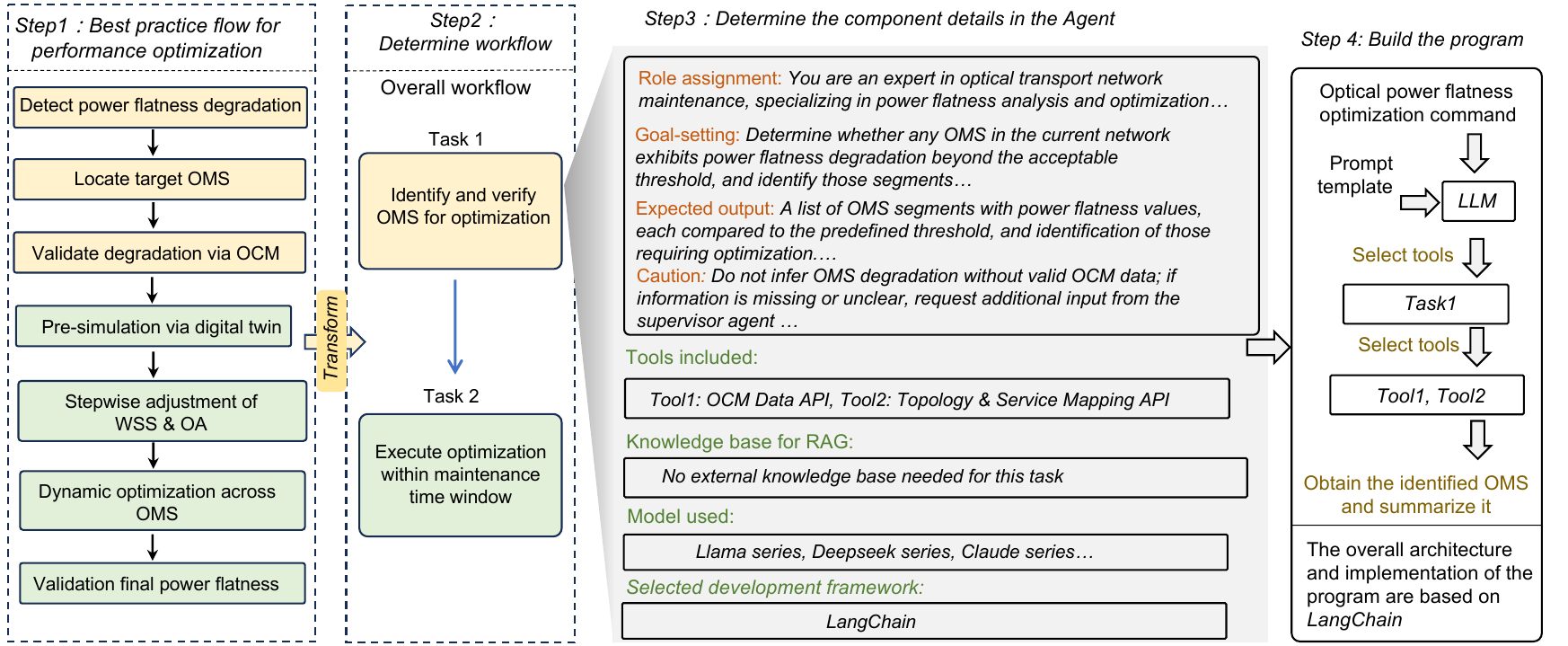}
\caption{Design and construction of the optical network performance optimization Agent for optical power flatness enhancement. The diagram presents the expert workflow for identifying and tuning OMS segments, along with corresponding task breakdowns, tool interfaces, prompt design, and Agent orchestration mechanisms.}
\label{Fig7}
\end{figure*}

We conducted an illustrative validation of the demand confirmation and resource selection subtasks shown in Fig. \ref{Fig6} using LangChain 0.2.16 and DeepSeek-V3.2-Exp. The constructed network topology, consisting of five nodes, is shown in Fig. \ref{Fig6B}. When a user requests to add four new channels between nodes A and C, the LLM invokes the RWA tool. In this process, we primarily consider three metrics: fiber routing latency, link stability score, and the number of available channels. The LLM evaluates all candidate fiber paths that satisfy the channel requirement and selects the optimal route A–B–C based on a composite score. For wavelength selection, the available idle channels are assigned sequentially, after which the QoT estimation tool is invoked to verify that the GSNR margins of the four channels on the A–B–C path all meet the required thresholds, thereby confirming this path as the final solution.

After route and spectrum selection is complete, the next task node—device configuration—must wait for the designated maintenance window and be triggered by the Supervisor Agent. The remaining two task nodes (device configuration deployment and performance validation/optimization) could follow a similar design pattern as the first, with careful tailoring of prompt templates, tools, and LLM models. Each node’s design should reflect its specific requirements for network performance and state data, while model selection must strike a balance between reasoning accuracy and computational cost.

It should be noted that subsequent tasks often involve live network configuration adjustments and the integration of various data and notification platforms, leading to higher task complexity and operational risk. To ensure accurate decision-making by the LLM, the prompt template should typically be upgraded from Standard Prompting to Chain-of-Thought Prompting. Moreover, the choice between Zero-Shot-CoT and Few-Shot-CoT should depend on network scale, task complexity, and empirical performance.

Additionally, the workflows for optical channel addition and deletion can be designed as parallel, independent pipelines, each with its own task nodes. This modular approach simplifies Agent implementation and improves maintainability.

\subsection{Agent for Performance Optimization}

\begin{figure*}[h]
\centering
\includegraphics[width=0.99\textwidth]{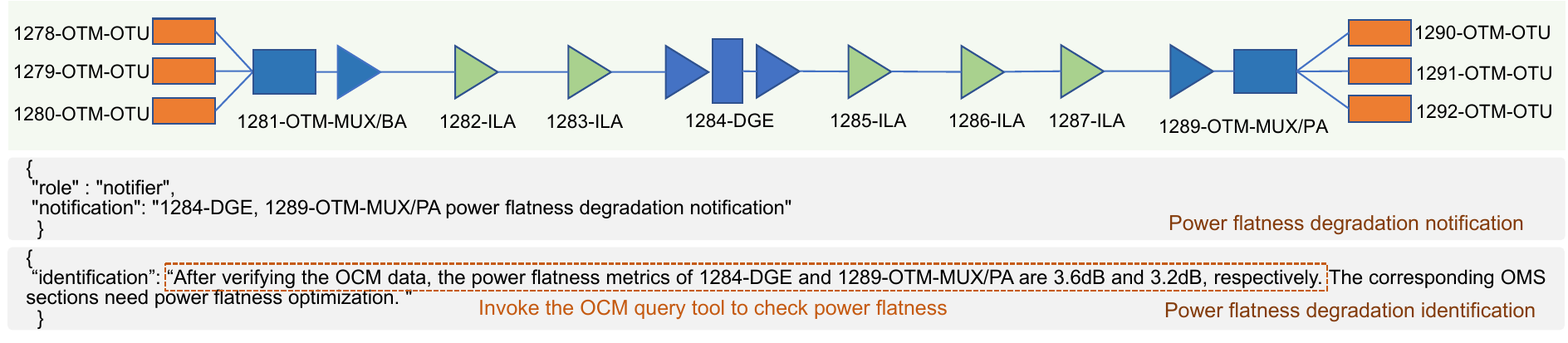}
\caption{Constructed multi-span transmission link and illustrative validation results for identifying OMS segments requiring power flatness optimizations.}
\label{Fig7B}
\end{figure*}

In optical networks, transmission quality fluctuations and degradation of inter-channel performance consistency frequently occur due to factors such as fiber degradation, equipment aging, and wavelength channel additions or deletions. To ensure network stability and reliability, timely performance optimization is essential. Among the various optimization goals, power flatness is the most commonly targeted, as engineers can directly monitor channel power through optical channel monitors (OCMs) and use it as feedback for adjustment. This makes implementation relatively feasible. However, when power flatness optimization is performed manually, it typically requires tuning each OMS individually, which is time-consuming and prone to configuration errors.

From a higher-level perspective, optimizing for both flat and maximized GSNR can bring the network’s transmission capacity close to its theoretical maximum. Under the condition of an accurate digital twin, optimal configurations can be identified through parameter tuning. The following section focuses on the task of power flatness optimization and outlines the design process for the performance optimization Agent.

The first step is to extract best practices from current human-in-the-loop power flatness optimization processes. When power flatness in a specific OMS falls below a defined threshold, network operators are notified either through routine inspections or via alarms proactively reported by devices equipped with OCMs. Operators then locate the affected OMS segment based on alarm information.

Since power flatness tuning involves adjusting the VOAs on WSS and the gain of optical amplifiers, it is considered a high-risk configuration change. Thus, it must be executed within authorized maintenance windows. In practice, engineers begin from the transmitting side of the OMS, fine-tuning VOAs incrementally and observing power flatness changes via OCM feedback in a closed-loop iterative manner. If a wavelength remains underpowered even when its VOA is set to the minimum, the OA gain needs to be increased while VOAs of other channels are slightly raised to maintain balance. Once an OMS segment is optimized, downstream segments must be checked for performance impacts. If flatness degrades beyond tolerable limits, further optimization is required downstream.

To ensure operational safety and effectiveness, a digital twin system can be used to simulate and preview configuration changes. This requires the digital twin to support fast updates and real-time interaction so that it can reflect the effect of each optimization step and guide the next. The optimization task is only considered complete once all affected OMS segments are adjusted and verified.

Based on the above, the performance optimization Agent’s workflow can be divided into two key tasks: (1) identifying the OMS segments that require optimization and verifying power flatness, and (2) executing the tuning process during the operational time window, as shown in Fig. \ref{Fig7}.

This task may be delegated by the Supervisor Agent or initiated independently by the performance optimization Agent, assuming the latter has a built-in power flatness inspection mechanism. Taking the latter case as an example, the first task involves several specific design elements. In the prompt template, the Agent should be assigned the role of a "power flatness optimization expert," and the value of the task and the method for calculating flatness should be clearly stated. The goal is to identify OMS segments with sub-threshold flatness and output the flatness values. To prevent hallucinations, the prompt should also instruct the LLM to avoid speculating on uncertain inputs.

The tools required for this task include interfaces to access OCM scan data across the network. With the obtained power spectrum, flatness can be easily computed by finding the difference between the maximum and minimum power values across channels, and compared against the predefined threshold. Additionally, the Agent may be equipped with APIs to query network topology and service connectivity, enabling spatial correlation analysis among affected OMS segments and identifying related service paths. Typically, this task does not require an external knowledge base. As the task primarily involves querying and logical reasoning, a lightweight LLM may be selected, based on test results. LangChain or other mainstream frameworks can be used to implement the Agent.

An illustrative validation result for the subtask of identifying OMS segments requiring power flatness optimization is presented in Fig. \ref{Fig7B}. A two-OMS multi-span transmission system was constructed. When the 1284-DGE and 1289-OTM-MUX/BA Agents reported flatness degradation, the LLM automatically obtained OCM measurement data from both sites via API access and, based on the actual flatness values, correctly identified the segments that required power flatness optimization.

For the second task of executing power flatness tuning, the overall workflow follows a similar design. However, special attention must be given to ensuring the safe execution of configuration changes. Through appropriate prompt design and digital twin API integration, the Agent should be capable of simulating the impact of adjusting WSS and OA parameters. Furthermore, it must adapt dynamically based on real-time network feedback after configuration commands are issued. This capability is critical to ensure network reliability throughout the tuning process.

\subsection{Agent for Optical Network Fault Management}

\begin{figure*}[htbp]
\centering
\includegraphics[width=0.99\textwidth]{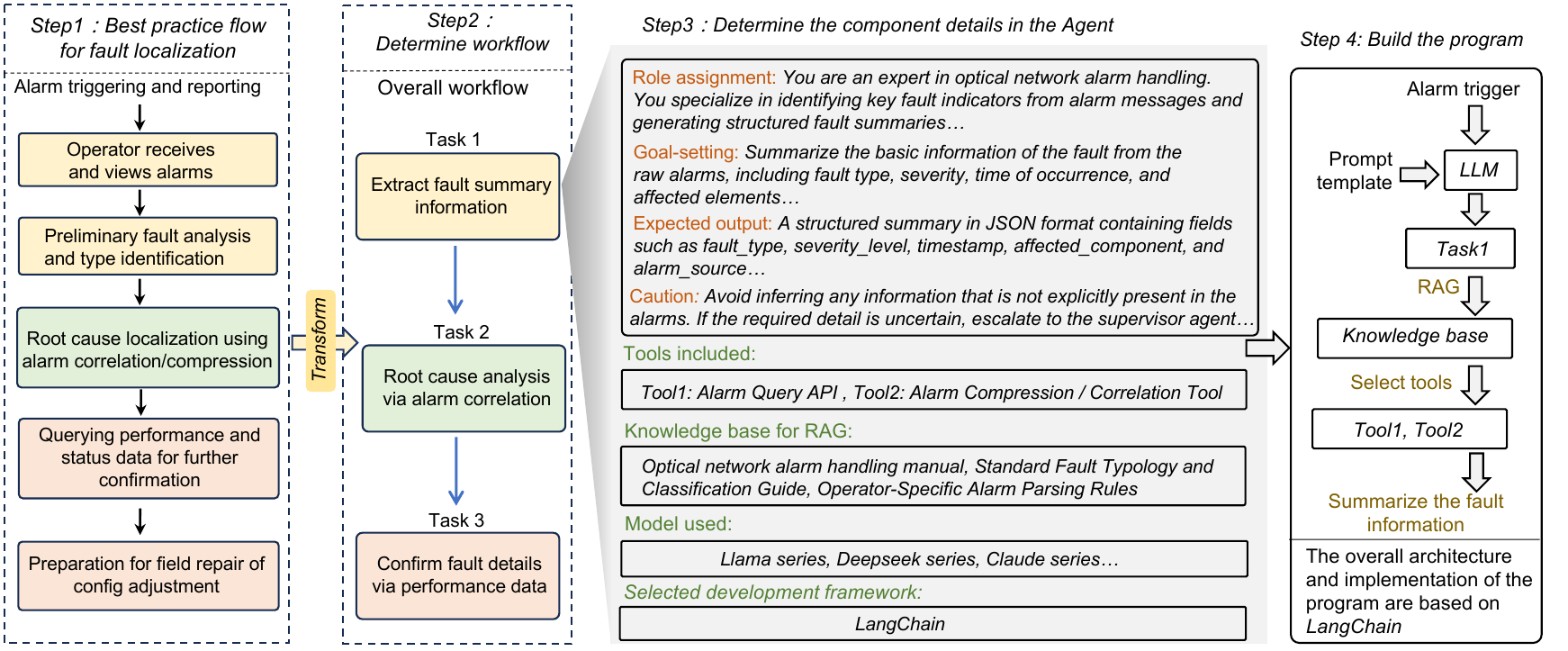}
\caption{Construction workflow of the fault management Agent for root cause localization. The figure outlines the operational best practices in live networks, maps them into an Agent-executable workflow, and provides detailed component design and development methodology for alarm parsing and fault analysis.}
\label{Fig8}
\end{figure*}

When introducing LLM-based AI Agents for optical network fault management, one of the primary objectives is to effectively manage the most frequent and impactful fault categories, namely fiber-related issues such as fiber cuts, jitter, and degradation, as well as equipment faults including board and module failures. In practice, fault management encompasses multiple distinct yet interrelated tasks, including fault prediction, anomaly detection, root cause localization, and recovery. Since these tasks are generally independent in time, a fault management Agent must support parallel workflows to handle them concurrently. This section focuses on the Agent design for fault localization, which remains central to the overall fault management process.

In current operational workflows, fault localization relies heavily on analyzing alarms and performance degradation symptoms. When a fault occurs, optical devices generate and report alarms based on predefined rules. For example, a fiber cut will typically trigger LOS-type alarms such as POWER\_LOS and OSC\_LOS at downstream optical layer boards, while service boards may report alarms like R\_LOS to indicate service loss. These alarms are aggregated and visualized through the NMS interface, forming the entry point for fault analysis. Network engineers then analyze the alarm patterns using predefined alarm correlations and device roles to determine the fault type, identify the affected subnet, and ideally pinpoint the specific fiber span, board, or module responsible for the fault. Alarm compression algorithms may also be used to accelerate the diagnostic process. Once the fault scope is narrowed, engineers consult historical performance metrics, such as received optical power levels, to assess the extent of fiber degradation and determine appropriate corrective actions, such as adjusting optical gain or initiating physical inspection.

The sequence of alarm reception, analysis, and verification is largely consistent across both fiber and equipment fault scenarios, providing a dependable foundation for designing an LLM-based Agent workflow. For fault localization, the workflow can be logically organized into three main phases: (1) initial extraction of fault information, (2) root cause inference based on alarms, and (3) verification using real-time or historical performance and status information, as shown in Fig. \ref{Fig8}. Although these steps are often blended in manual troubleshooting, explicit separation into structured task stages is essential for automated execution by an Agent.

In the first phase, the Agent processes incoming alarm data to extract high-level fault descriptors such as event timestamp, severity, and affected network domains. While often implicit in manual workflows, this stage is critical for triggering the subsequent steps in the Agent pipeline. The second phase involves causal analysis over alarm patterns, exploiting spatial and temporal correlations to infer root causes. Finally, the third phase gathers relevant device metrics to confirm or refine the diagnosis, ensuring sufficient evidence before recommending actions.

Each phase can be supported by tailored prompts and external tools. For instance, in the first phase, a well-designed prompt can instruct the LLM to act as an "optical network alarm analysis expert," with a clearly defined goal: extract fault descriptors in a structured format from raw alarm logs. To ensure consistency and minimize hallucination risk, the prompt should specify output templates and handling strategies for uncertain or missing information. Tool-wise, the Agent may rely on alarm query APIs, log parsers, or alarm correlation engines. Knowledge retrieval can be enhanced via RAG by supplying fault type handbooks and rule-based classification guides as external resources. Given the moderate complexity of this step, a lightweight LLM is generally sufficient and can be efficiently deployed using Agent orchestration frameworks such as LangChain.

\begin{figure}[!t]
\centering
\includegraphics[width=\columnwidth]{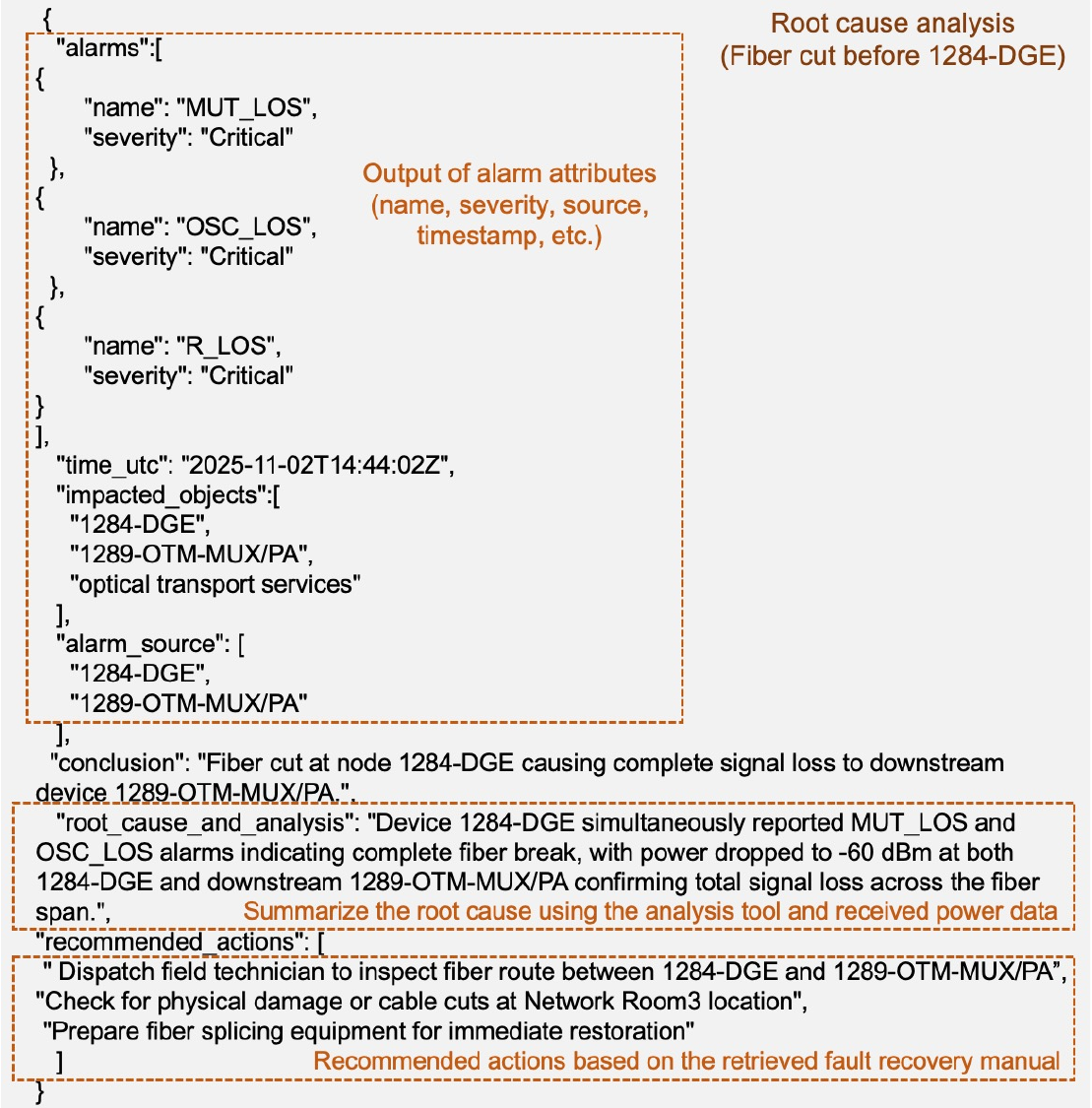}
\caption{Illustrative validation results of root cause analysis based on alarm information.}
\label{Fig8B}
\end{figure}

For the fault management scenario, an illustrative validation of root cause analysis was conducted using the transmission system shown in Fig. \ref{Fig7B}. A fiber break was simulated between the 1283-ILA and 1284-DGE nodes, resulting in MUT\_LOS and OSC\_LOS alarms reported by 1284-DGE, as well as an R\_LOS alarm reported by 1289-OTM-MUX/PA. After extracting key information such as alarm names, severity, sources, and timestamps, the LLM invoked the fault root cause analysis tool and successfully identified the fiber break as the root cause. Furthermore, since a troubleshooting handbook was provided, the LLM was also able to generate appropriate fault resolution suggestions, as shown in Fig. \ref{Fig8B}.

Subsequent stages, particularly root cause analysis and performance-based confirmation, require the Agent to interpret network topology, understand service mappings, and incorporate historical behavior patterns. Tools for topology traversal, configuration inspection, and time-series metric analysis become essential here. With a multi-modal data processing pipeline and access to rich monitoring interfaces, the fault localization Agent can reproduce the diagnostic logic of expert engineers, achieving both efficiency and accuracy in high-volume environments.

Other fault management tasks, such as proactive fault prediction and real-time anomaly detection, can be developed following similar design principles. For each task, the Agent's workflow should be aligned with operational best practices, clearly scoped into manageable task nodes, and supported by appropriate prompts, tools, and domain-specific resources. As automation matures within optical networks, fault management Agents can be progressively expanded in capability and precision, laying a robust foundation for intelligent, self-sustaining network operations.

\section{Challenges}

Although the proposed multi-Agent collaborative O\&M architecture is designed with consideration of practical network conditions and addresses multiple categories of tasks with executable workflows, the deployment of LLM-based AI Agents in real-world optical networks still faces several key challenges.

\subsection{Real-time Data Access}

The first challenge lies in the real-time acquisition of operational data from live optical networks. In most high-value O\&M scenarios, higher-level automation depends on continuous and timely awareness of network performance and state information. For example, when performing optical power equalization in large-scale systems, parameter adjustments in one optical multiplex section (OMS) often introduce cascading effects on multiple downstream OMS segments. To ensure operational safety, each adjustment must be applied in small steps and verified through pre-simulation using a digital twin. However, the efficiency of such simulations is largely constrained by the data acquisition rate—if network data cannot be retrieved in real time, the overall optimization loop becomes significantly slower.

Similarly, in fault prediction tasks, fine-grained operational data at the millisecond level are essential for capturing early anomaly patterns. Yet, most optical networks currently deployed by operators still rely on historical data with coarse time granularity (typically every 15 minutes), which cannot meet the real-time requirements. Moreover, frequent high-rate API requests to collect real-time data may cause device response delays or even impact network stability.

To align with an LLM-driven O\&M architecture, the data-pushing mechanisms of optical network elements must be further optimized. In addition, the appropriate preprocessing strategies for different types of raw monitoring data before they are fed to the LLM remain an open research topic, as these directly affect the model’s analytical accuracy and task execution efficiency. Overall, enhancing the real-time data acquisition and preprocessing pipeline is a prerequisite for enabling higher levels of automation in optical network operations.

\subsection{Digital Twin Modeling Capabilities}

The second challenge lies in building high-fidelity digital twins for large-scale optical networks. Digital twins have become a fundamental enabler for key O\&M scenarios such as operation rehearsal, performance evaluation, and fault prediction. Their modeling accuracy and computational efficiency directly determine the practical boundary of applying LLMs for intelligent optical network management.

However, in current O\&M practices, the precision and scalability of digital twin models are still limited by several factors, including inaccurate fiber parameters, insufficient amplifier modeling fidelity, and the inherent complexity of large-scale network topologies and service paths. Consequently, there remains a noticeable gap between simulation accuracy and deployment efficiency. To bridge this gap, further advancements are needed in several directions, such as physical parameter calibration, parallel simulation scheduling, and data-driven model construction.

In addition, most existing academic and industrial efforts have focused primarily on digital twins at the physical layer of optical transmission systems, while the modeling of higher-layer network properties—such as network capacity, latency, and service-level agreement (SLA) requirements—remains underdeveloped. Therefore, achieving high-fidelity, multi-layer digital twins continues to be an open challenge and is essential for providing reliable support to LLM-driven autonomous O\&M in optical networks.

\subsection{LLM Reliability}

The third challenge concerns the reliability and safety of LLMs when applied to optical network O\&M. As the foundational infrastructure supporting upper-layer services, optical networks require extremely high stability and operational accuracy. Introducing LLMs into such environments therefore raises a critical question: how to ensure that the model’s analyses and decisions are logically sound and will not induce unintended network disruptions.

At the current stage of development, hallucination remains an inherent limitation of LLMs, and fully eliminating it is not yet feasible. Existing studies show that techniques such as CoT prompting and RAG can mitigate hallucinations to some extent, especially when handling complex tasks or domain-specific reasoning. However, their effectiveness is domain-dependent, and no comprehensive methodology has been established to guarantee hallucination-free behavior.

To prevent unreliable model outputs from posing operational risks to live networks, additional safeguard mechanisms must be incorporated between the LLM’s decision and the actual execution actions. For example, before applying configuration changes to network devices, the proposed framework requires that each action be pre-validated through a digital twin to ensure that the impact remains within acceptable bounds. Furthermore, for high-risk operations, a mandatory human approval step should be enforced to maintain strict control over operational safety.

Overall, the reliability and safety of LLM-driven decisions must be carefully addressed to prevent inappropriate actions or hallucinated outputs from affecting service stability in optical networks.

\section{Conclusions}

This paper starts from the practical needs of optical network operations and systematically examines the potential and applicable scenarios of LLMs for enhancing intelligent network management. A general multi-Agent collaborative architecture is proposed, in which a Supervisor Agent orchestrates task decomposition and the coordination of sub-Agents. By integrating prompt engineering, tool invocation, and workflow design, the architecture delivers Agent-based solutions for key tasks such as channel provisioning, performance optimization, and fault management. Furthermore, this paper discuss the key challenges faced in deploying LLMs in real-world networks, particularly in terms of real-time data access, DT modeling capabilities, and model reliability. Future work will focus on validating the proposed framework through phased deployment in operational networks, and further exploring its integration with existing NMS and SDN controller to enable seamless orchestration. Overall, this work provides a conceptual framework for applying LLMs to optical networks and lays the foundation for building a closed-loop, autonomous, and intelligent O\&M system at scale.

\section*{Funding}National Natural Science Foundation of China (U24B20133, 62522104).

% Bibliography
\bibliography{sample}

% Manual citation list
% \begin{thebibliography}{1}
% \bibitem{Zhang:14}
% Y.~Zhang, S.~Qiao, L.~Sun, Q.~W. Shi, W.~Huang, %L.~Li, and Z.~Yang,
 % \enquote{Photoinduced active terahertz metamaterials with nanostructured
  % vanadium dioxide film deposited by sol-gel method,} Opt. Express \textbf{22},
  % 11070--11078 (2014).
% \end{thebibliography}

\end{document}